\documentclass[twocolumn,noshowpacs,amssymb,amsfonts,letterpaper,nofootinbib,aps,prd]{revtex4}

\usepackage{amsmath,graphicx}

\def\VEV#1{\left\langle #1 \right\rangle}

\begin{document} 

\title{Correlation of inflation-produced magnetic fields with
scalar fluctuations}
\author{Robert R. Caldwell}
\author{Leonardo Motta}\email{leonardo.motta@dartmouth.edu}
\affiliation{Department of Physics \& Astronomy, HB 6127 Wilder
     Lab, Dartmouth College, Hanover, NH 03755, USA} 
\author{Marc Kamionkowski}
\affiliation{California Institute of Technology, Mail Code
     350-17, Pasadena, California 91125, USA} 
\affiliation{Dept. of Physics \& Astronomy, 
	Johns Hopkins University, Baltimore, MD 21218, USA} 

\date{\today}

\keywords{magnetic field, primordial; magnetic field, galactic; inflation; dynamos; in-in formalism}

\begin{abstract}
If the conformal invariance of electromagnetism is broken during
inflation, then primordial magnetic fields may be produced. If
this symmetry breaking is generated by the coupling between
electromagnetism and a scalar field---e.g. the inflaton,
curvaton, or the Ricci scalar---then these magnetic fields may
be correlated with primordial density perturbations, opening a
new window to the study of non-gaussianity in
cosmology. In order to illustrate, we couple
electromagnetism to an auxiliary scalar field in a de Sitter
background.  We calculate the power spectra for scalar-field
perturbations and magnetic fields, showing how a scale-free magnetic field spectrum with
rms amplitude of $\sim {\rm nG}$ at Mpc scales may be achieved.
We explore the Fourier-space dependence of the
cross-correlation between the scalar field and magnetic fields, showing that the dimensionless amplitude,
measured in units of the power spectra, can grow as large as
$\sim 500 H_I/M$, where $H_I$ is the inflationary Hubble
constant and $M$ is the effective mass scale of the coupling.
\end{abstract}

\maketitle

\section{\label{sec:intro}Introduction}

The predictions of the simplest single-field slow-roll models of
inflation agree remarkably well with current cosmological data,
yet experience gained from effective field theories suggests
that this model is likely not the whole story.   A vast
literature has now arisen to explore ultraviolet completions 
and their predictions for future, more sensitive, observations
\cite{Weinberg:2008hq,Senatore:2010wk}. One of the principle
lines of investigation has been the predictions for
non-gaussianity due to self-couplings, nontrivial inflaton
kinetic terms or interactions between multiple fields associated
with inflation \cite{Bartolo:2004if,Komatsu:2009kd,Huterer:2010en}.

Another possibility for beyond single field slow roll physics is
coupling of the inflaton, or some other spectator field, to
electromagnetism. If such a coupling breaks the conformal
invariance of electromagnetism, then quantum fluctuations in the
electromagnetic field may be amplified into classical magnetic
fields in much the same way as quantum flucutations in the
inflaton (graviton) become density perturbations (gravitational
waves). It has been suggested that such inflation-produced
magnetic fields may provide the seed fields required for
galactic
dynamos~\cite{Turner:1987bw,Ratra:1991bn,Widrow:2002ud,Bamba:2003av,Kunze:2007ph,Campanelli:2008qp,Demozzi:2009fu,Kunze:2009bs,Kandus:2010nw},
but it may also be that the signatures of such magnetic fields
may be observed in the cosmic microwave
background~\cite{Lewis:2004kg,Kosowsky:1996yc,Kosowsky:2004zh,Giovannini:2008aa,Kahniashvili:2008hx,Kristiansen:2008tx,Seshadri:2009sy,Caprini:2009vk,Cai:2010uw,Shiraishi:2010kd,Brown:2010jd,Giovannini:2009zq,Yamazaki:2010nf,Kahniashvili:2010wm},
and thus shed light on inflation, even if they are unrelated to
galactic magnetism.  Either way, the search for primordial
magnetic fields provides an additional observational probe of
the physics of inflation to parallel that obtained from
non-gaussianity searches.

Here we explore the cross-correlation between primordial
magnetic fields and a scalar field in a toy model in which the
scalar field is coupled to electromagnetism, with no gravity, in
a fixed de Sitter background.  The homogeneous time evolution of
the scalar field breaks the conformal invariance of
electromagnetism.  We first calculate the quantum mechanical
spectrum of scalar- and magnetic-field fluctuations produced,
and we then calculate the cross-correlation between the scalar
and magnetic fields.

If the scalar field is a curvaton field, and if that curvaton is
responsible for primordial perturbations, then the
scalar-field--magnetic-field cross correlation we calculate will
be precisely the density--magnetic-field correlation observed in
the Universe today. Our calculation also illustrates the
principal ingredients that will arise in a
density-perturbation--magnetic-field correlation if the scalar
field is the inflaton.

In Section~\ref{sec:toymodel} we introduce our model, work out
the dynamical behavior, and evaluate the two-point statistics of
the scalar and magnetic fields. In Section~\ref{sec:corr} we
present the calculation of the cross-correlation, and we analyze
its behavior in Sec.~\ref{sec:analysis}.  We conclude in
Sec.~\ref{sec:conclusion}. Throughout, we work in spatially-flat
Robertson-Walker coordinates, with line-element $ds^2 =
a^2(\eta)(-d\eta^2 + d\vec x^2)$.

\section{\label{sec:toymodel}Mechanism of Magnetic Field Amplification}

The action for our model is
\begin{equation}
     S=\int d^4 x \; \sqrt{-g}\left( -\frac{1}{4}W(\phi)
     F_{\mu\nu}F^{\mu\nu}-\frac{1}{2} (\partial\phi)^2
     - V(\phi) \right), \nonumber
\end{equation}
where $\phi(\vec x,t)$ is the scalar field, and $F_{\mu\nu}$ the
electromagnetic field-strength tensor.  The scalar-field
potential is $V(\phi) = -3 n M H_I^2 \phi$, and the coupling
function is $W(\phi) = e^{2 \phi/M}$.  We suppose that some
other field is driving inflation.  In practice we consider a
fixed de Sitter background with Hubble constant $H_I$, whereby
the scale factor is $a(\eta) = -1/(\eta H_I)$ for the run of
conformal times $-\infty < \eta \le \eta_I <0$ and $\eta_I$
marks the end of inflation.  Aspects of this model have
previously been studied \cite{Bamba:2003av,Demozzi:2009fu}, but
we revisit the details in preparation for our later
calculations.

\subsection{Scalar Field}

The scalar-field equation of motion is
\begin{equation}
     \Box\phi = \frac{\partial V}{\partial \phi} + \frac{1}{4}
     \frac{\partial W}{\partial \phi} F_{\mu\nu} F^{\mu\nu},
\end{equation}
and it has a solution,
\begin{equation}
     \phi = c_0 + c_1 \eta^3 - n M \ln(\eta/\eta_I),
\end{equation}
where we assume there is no homogeneous electric or magnetic
field. We take the integration constants $c_0$ and $c_1$ to
vanish so that $W(\phi)=1$ at the end of inflation.  In this
way, the usual electromagnetic Lagrangian is recovered for the
post-inflationary epoch, and we assume that some mechanism stops
the  subsequent evolution of $\phi$, so that the standard
Maxwell equations are preserved at all times after inflation. We
also define $I(\eta) \equiv \left[W\left(\phi(\eta)\right)
\right]^{1/2} = \left({\eta}/{\eta_I} \right)^{-n}$, which will
appear in our analysis of the electromagnetic field.

The scalar field has fluctuations $\delta\phi(\vec x,\eta)$,
about its homogeneous component, described by the evolution
equation,
\begin{equation}
     \delta\phi'' + 2 {\cal H}\delta\phi' + (a^2 V_{\phi\phi} -
     \nabla^2) \delta\phi = 0,
\end{equation}
where ${\cal H}=a'/a$ and $\nabla^2$ is the spatial Laplace
operator. To be clear, we fix the background to be pure de
Sitter spacetime and subsequently ``turn off'' gravity, so that
there are no fluctuations of the spacetime metric. Following
standard procedures, the quantized scalar field is decomposed in
terms of time-dependent mode functions $\delta\phi_k$,
\begin{equation}\label{eq:dphi}
     \delta\phi(\vec x, \eta) = \int \frac{ d^3 k}{(2\pi)^3}
     \left[ e^{i \vec k \cdot \vec x} \delta\phi_k(\eta)
     \hat a_k + {\rm h.c.}\right] ,
\end{equation}
where $\hat a_k$ and $\hat a_k^\dagger$ are respectively
annihilation and creation operators that satisfy $[\hat a_k,
\hat a_{k'}^\dagger] = (2\pi)^3 \delta(\vec k - \vec k')$. 
The uncertainty relation for the  scalar field and its conjugate
momentum $\delta\phi'$,
\begin{equation}
     [\delta\phi(\vec x,\eta), \delta\phi'(\vec y, \eta)] = i
     \delta(\vec x - \vec y)/ a^2(\eta),
\end{equation}
results in a constraint to the two linearly independent
solutions to the mode equation.  Because the effective mass is
zero, $V_{\phi\phi}=0$, we obtain the solution,
\begin{equation}
     \delta\phi_k(\eta) = \frac{H_I}{\sqrt{2}}
     \frac{i-k\eta}{k^{3/2}} e^{-i k \eta},
\end{equation}
corresponding to the Bunch-Davies state, having positive
frequency in the remote past, $\eta \to -\infty$ for $k |\eta |
\gg 1$. The requirement $\delta\phi \ll \phi$ that the
fluctuations are small translates into the bound $H_I/M \ll 1$.
Finally, the two-point correlation function is
\begin{equation}
     \VEV{ \delta\phi(\vec x,\eta)\delta\phi(\vec y, \eta)} =
     \int \frac{d^3 k}{(2 \pi)^3}
     e^{i\vec k \cdot (\vec x -\vec y)}
     P_{\delta\phi}(k)
\end{equation}
where the scalar-field power spectrum---defined by
$\VEV{\delta\phi_{\vec k}  \delta\phi^{*}_{\vec k'}} = (2\pi)^3
\delta_D(\vec k-\vec k') P_{\delta\phi}(k)$ and $\delta_D$ is
the Dirac delta function---is $P_{\delta\phi}(k) = H_I^2/2
k^3$, valid for modes outside the horizon at the end of
inflation.
  
The root-mean-squared amplitude---the correlation function at
zero lag (at $\vec x=\vec y$)---is divergent at both the
infrared and ultraviolet limits. Hence, we bound the run of
wavenumbers to $[k_\text{min},\, k_\text{max}]$, so that
\begin{equation}
     \delta\phi_{\text{rms}} \equiv \VEV{ (\delta\phi)^2}^{1/2} =
     \frac{H_I}{2 \pi} \left(\ln k_\text{max}/k_\text{min}
     \right)^{\frac{1}{2}}
\end{equation}
gives the rms scalar-field fluctuation. In practice, we
associate the minimum wavenumber with the present-day Hubble
radius---i.e., $k_\text{min} = 2 \pi H_0$---and the maximum
wavenumber with an astrophysical scale that we indicate by
$\lambda$.

\subsection{Electromagnetism}

The full action for electromagnetism includes not only the free
Maxwell field, but also the coupling to charged particles as
well as the action for the charged particles themselves.
Including these additional terms, we may write
\begin{equation}
\label{eq:origaction}
     S_{\rm em} = -\int d^4 x \; \sqrt{-g} \left[
     \frac{1}{4}I^2(\phi) F_{\mu\nu}F^{\mu\nu} + A^\mu J_\mu +
     {\cal L}_q \right],
\end{equation}
where ${\cal L}_q$ is the Lagrangian for charged particles. The
electromagnetic coupling, or electric charge, is inversely
proportional to $I(\eta)$. Consequently, in the case $n>0$ the
coupling is strong at early times \cite{Demozzi:2009fu}.  Such a
strong-coupling scenario has previously been dismissed
\cite{Demozzi:2009fu}, since the free-field behavior of
electromagnetic waves would no longer be valid.  We therefore
consider here the alternative Lagrangian,
\begin{equation}
\label{eq:newaction}
     S_{\rm em} = -\int  d^4 x \; \sqrt{-g} \, I^2(\phi)\left[
     \frac{1}{4} F_{\mu\nu}F^{\mu\nu} + A^\mu J_\mu + {\cal
     L}_q\right],
\end{equation}
in which the conformal factor $I^2(\phi)$ is moved outside the
entire electromagnetic-sector Lagrangian.  With this
modification, the  strong-coupling problem is alleviated. This
Lagrangian could arise if $\phi$ is a dilaton field, although in
that case we would not expect the conformal factor $I^2(\phi)$
to also multiply the mass term of the charged particle.

The effect of the coupling function on Maxwell's equations is
straightforward. In the absence of charges or currents, only
Ampere's equation is modified, to
\begin{equation}
     \vec\nabla \times \vec B = \frac{1}{a^2 W}
     \frac{\partial}{\partial \eta} \left( a^2 W \vec E \right),
\end{equation} 
where we have assumed $W$ is solely a function of (conformal)
time.  Faraday's law remains unaltered:
\begin{equation}
     \vec\nabla \times \vec E = - \frac{1}{a^2}
     \frac{\partial}{\partial \eta} \left( a^2 \vec E \right).
\end{equation}
A homogeneous magnetic field therefore scales with conformal time as
$|\vec B| \propto \eta^2$, while a homogeneous electric field
scales as $|\vec E| \propto \eta^{2+2n}$.  The magnetic- and
electric-field energy densities scale as $\rho_B =W B^2/8\pi
\propto \eta^{4+2n}$ and $\rho_E =WE^2/8\pi \propto
\eta^{4+2n}$.  Thus, for $n=2$ and $n=-3$ (special cases we will
consider below), the magnetic-field energy density remains
constant or decays, respectively.  The electric-field energy
density decays for $n=2$, but it grows, as $\rho_E \propto
\eta^{-2}$, for $n=-3$.  In this latter case, the energy density in
the electric-field component of the quantum-mechanically induced
electromagnetic fields will, if inflation goes on long enough,
ultimately dominate the energy density, $\sim 3 H_I^2/(8\pi G)$,
in the inflaton.  As we will see below (see also
Ref.~\cite{Demozzi:2009fu}), this then severely restricts the
number of $e$-foldings of inflation.  We will thus ultimately
discard the $n=-3$ case.

\subsection{Quantum Fluctuations of the Magnetic Field}

The action for the free field theory is
\begin{eqnarray}
     S_{\rm em} &=& -\int  d^4 x \; \sqrt{-g} \,
     \frac{1}{4}I^2(\phi) F_{\mu\nu}F^{\mu\nu} \nonumber \\
     &=& \int d\eta\,
     d^3 x \,  \left[I(\eta) \right]^2 \left( \frac{1}{2} A_i'^2
      -  \frac{1}{4}(\partial_i A_j - \partial_j A_i)^2 \right),\nonumber \\
\end{eqnarray}
in the Coulomb gauge, where $A_i$ is the vector potential. The
Latin indices here are contracted using the spatial part of the
Minkowski metric. Defining the vector field $V_i = I(\eta) A_i$
we can bring the kinetic term to canonical form, whereby
\begin{equation}
     S_{\rm em} = \int d\eta \, d^3 x \, \frac{1}{2}\left[
     V_i'^2 - (\partial_i V_j)^2 + \frac{I''}{I}V_i^2\right],
\end{equation}
after some integrations by parts. The quantized field $V_i$ is
expanded in terms of time-dependent mode functions $v_k(\eta)$,
\begin{equation}\label{eq:Vi}
     V_i(\vec k, \eta) = \sum_{\sigma=1}^2 \int \frac{d^3 k}{(2\pi)^3} \left[
      e^{i \vec k \cdot \vec x} v_k(\eta)
     e_i^{(\sigma)}(\hat k)\hat b_\sigma(k) + {\rm h.c.}\right],
\end{equation}
where $\hat b, \hat b^\dagger$ are annihilation and creation
operators satisfying $[\hat b_\sigma(\vec k), \hat
b^\dagger_{\sigma'}(\vec k)] = (2\pi)^3 \delta_{\sigma,\sigma'}
\delta_D(\vec k - \vec k')$, where $e_i^{(\sigma)}$ is the
polarization vector, $\sigma$ sums over the two
linear-polarization states, and $\sum_\sigma e_i^{(\sigma)}(\hat
k) e_j^{(\sigma)}(\hat k) = \delta_{ij}-\hat k_i \hat k_j$ which
further ensures transversality as a consequence of the gauge choice.
Canonical quantization means that the vector field and its
conjugate momentum $V_i'$ satisfy the commutation relation,
\begin{equation}
     [V_i(\vec x,\eta), V_j'(\vec y, \eta)] = i \delta_{ij}
     \delta(\vec x - \vec y),
\end{equation}
which results in a constraint to the two linearly independent
solutions to the mode equation. The scalar field contributes an
effective time-dependent mass term to the vector field, so that
the mode functions obey the equation,
\begin{equation}
     v_k'' + \left(k^2 - \frac{I''}{I} \right)v_k = 0,
\end{equation}
where ${I''}/{I} = n(n+1)/\eta^2$, is positive for $n>0$ or $n'
= n+1<0$. At high frequencies, $k|\eta| \gg 1$, the solutions are
oscillatory, but at low frequencies the scalar field causes
solutions to grow as $v_k \propto |\eta|^{-n},\, |\eta|^{1+n}$.
The normalized solution, having positive frequency in the remote
past, $\eta \to -\infty$, for $k |\eta | \gg 1$,  is
\begin{equation}
     v_k(\eta) =\sqrt{\frac{\pi}{2}} \frac{(-k \eta)^{1/2}}{\sqrt{2 k}} e^{i
     \pi(1+n)/2}H^{(1)}_{\frac{1}{2}+n}(-k\eta),
\end{equation}
where $H_n(x)$ is a Hankel function.  In this case, the
two-point correlation function for the magnetic field is 
\begin{eqnarray}
     \langle \vec B(\vec x,\eta)\cdot \vec B(\vec y,
     \eta)\rangle 
     &=& \frac{1}{a(\eta)^4} \left(\delta_{ij}
     \frac{\partial^2}{\partial x^k \partial y^k} -
     \frac{\partial^2}{\partial x^j \partial y^i}\right)
     \nonumber \\
     & & \times \VEV{A_i(\vec x,\eta)A_j(\vec y,\eta)} \nonumber \\ 
     &=&
      \int \frac{d^3 k}{(2 \pi)^3}
       e^{i\vec k \cdot (\vec x - \vec y)}
      P_B(k),
\end{eqnarray}
where 
\begin{equation}
\label{eq:fullpb}
     P_B(k) =  \frac{\pi}{2} \frac{H_I^4}{k^3}
     \left(\frac{\eta}{\eta_I}\right)^{2n}(-k \eta)^5
     H^{(1)}_{\frac{1}{2}+n}(-k\eta)
     H^{(2)}_{\frac{1}{2}+n}(-k\eta),
\end{equation}
is the magnetic-field power spectrum.  In the unamplified case,
corresponding to $n=0$, we have $P_B^{(0)} = k(H_I \eta_I)^4$ at
the end of inflation; the correlations in this case are then the
usual vacuum-fluctuation correlations.  Production of classical
long-wavelength magnetic fields occurs for $n>0$ or for $n' =
n+1 < 0$.  To treat both cases with a single expression, we
define $n_B = 4-2n$ for the case $n \ge 0$ and $n_B = 4+2 n'$
for $n' = n+1 < 0$. Consequently, the power spectrum is
\begin{equation}
\label{eq:pb}
     P_B \simeq  \frac{\Gamma(\frac{5-n_B}{2})^2}{\pi}\left({-k
     \eta_I / 2}\right)^{n_B -4} P_B^{(0)},
\end{equation}
for modes outside the horizon at the end of inflation. Since $k
|\eta_I | \ll 1$ for modes outside the horizon at the end of
inflation, and since $n_B-4<0$, the amplified ratio $P_B /
P_B^{(0)}$ can grow quite large.

The mean-squared magnetic-field power in long-wavelength modes
at the end of inflation, per logarithmic interval, is 
\begin{equation}
     \frac{d}{d\ln k} \langle B^2 \rangle \simeq
     \left(\frac{2}{\pi}\right)^3  \Gamma \left(\frac{5-n_B}{2}
     \right)^2   H_I^4 (-k \eta_I/2)^{n_B}.
\end{equation}
A scale-free spectrum $n_B=0$ can be achieved for $n=2,\,-3$.
Using $({\rm Gauss})^2/8 \pi = 1.91 \times
10^{-40}\,\rm{GeV}^4$, ${\rm Mpc} = 1.56 \times 10^{38}\,{\rm
GeV}^{-1}$, estimating $|\eta_I| \sim 10^{-27}\,{\rm Gpc}$
(consistent with $H_I \simeq 10^{14}$~GeV and $z_I \simeq
10^{28}$ for the redshift to the end of inflation), and then
redshifting to the present day, we find
\begin{equation}
     \frac{d}{d\ln k} \langle B^2 \rangle \simeq 10^{-18-24.3
     n_B} \frac{\Gamma(\frac{5-n_B}{2})^2}{\Gamma(5/2)^2} \left(
     \frac{k}{{\rm Mpc}^{-1}}\right)^{n_B}\, {\rm G}^2.
     \label{eqn:Bmag}
\end{equation}
If $n_B=0$ or $n=2$ or $-3$, then the field strength is roughly
$10^{-9}\,{\rm G}$, which may be sufficient to explain the
observed astrophysical and cosmological magnetic fields
\cite{Widrow:2002ud}.

\subsection{Energy Density of the Magnetic and Electric Fluctuations}

The same magnetic-field spectrum is obtained for two values of
the index $n$. However, the time evolution of the coupling
function $I(\eta)$ breaks the usual duality between electric and
magnetic fields, and the electric-field energy density may in
some cases increase, as discussed above.  We require the energy
density in superhorizon modes of the electromagnetic fields to
be smaller than the energy density of the inflaton, and thereby
derive now a restriction on the allowed values of $n$ and $H_I$.

The stress-energy tensor that appears as a source for the
Einstein equations is
\begin{equation}
     T^{\mu\nu} = I^2(\phi) \left( g_{\alpha\beta} F^{\mu\alpha}
     F^{\nu\beta}  - \frac{1}{4} g^{\mu\nu} F_{\alpha\beta}
     F^{\alpha\beta}\right).
\end{equation}
The energy density observed in the cosmic rest frame is
\begin{eqnarray}
     \rho_{EB} &=& \frac{I^2(\eta)}{2 a^4(\eta)}\langle A'_i
     A'_i + (\partial_i A_j ) (\partial_i A_j) -( \partial_i
     A_j ) (\partial_j A_i)\rangle  \cr
     &=& \lim_{\vec x \to \vec y}\int \frac{d^3 k}{(2 \pi)^3}
      e^{i\vec k \cdot( \vec x - \vec y)} P_{\rho}(k),
\end{eqnarray}
where the final term in the top line ultimately vanishes due to
the transversality condition $\hat k \cdot \vec e$. The
energy-density power spectrum consists of two terms, a kinetic
term due to the electric field and a spatial-gradient term due
to the magnetic field,
\begin{equation}
     P_\rho(k) =  \frac{3(2 \pi)^3}{2 a^4(\eta)} \left( \left|
     I(\eta) \left(\frac{ v_k(\eta)}{I(\eta)} \right)'
     \right|^2 + k^2  \left| v_k(\eta)
     \right|^2 \right).
\end{equation}
The integral over wavenumbers runs from $k_\text{min} =
-1/\eta_S$ to $k_\text{max} = -1/\eta$,  where $\eta_S = \eta_I
e^{N_I} $ is the conformal time at the beginning of $N_I$
$e$-foldings of inflation, thereby spanning the range of
wavelengths that have exited the horizon by the time $\eta$. The
pattern of behavior distinguishes two regimes,
\begin{equation}
     \rho_{EB} = H_I^4 \times
     \begin{cases}
     {\cal O}(1), &{\rm for}\quad  |n| \le 2, \cr
     {\cal O}(1) \times
     \left(\frac{\eta_S}{\eta}\right)^{2(|n|-1)}, &{\rm for}\quad |n| > 2. \cr
\end{cases}
\end{equation}
In the first case, which includes the scale-free solution
$n=2$, the energy density is simply proportional to $H_I^4$
which is always subdominant to the inflaton energy density.
However, the second case, which includes the other
scale-free solution $n=-3$, places severe restrictions, 
\begin{equation}
 |n| < 2 + \frac{1}{N_I} \ln \frac{M_P}{H_I},
\end{equation}
on the index $n$. Since observational constraints limit $H_I
\lesssim 10^{-5} M_P$, then to achieve at least 60 $e$-foldings of
inflation, the index is bounded by $|n| < 2.2$, thereby
eliminating the case $n=-3$. At the value $n=-2.2$,
Eq.~(\ref{eqn:Bmag}) tells us  that the magnetic-field strength
on Mpc scales is roughly $10^{-30}$G.  The case $n=2$, however,
safely satisfies the above bound and yields a nG magnetic
field as we have shown.

\section{Correlation of Magnetic Fields and Scalar Fluctuations}
\label{sec:corr}

We now evaluate the $(\delta\phi)BB$ correlation making use of
the in-in formalism \cite{Weinberg:2005vy}. After splitting the
Hamiltonian into a free part plus an interaction part $\hat
H_{\rm int}$, we may evaluate, to first order in perturbation
theory,
\begin{widetext}
\begin{equation}\label{eq:zvvtree}
     \VEV{ \frac{\delta\phi}{M} (\vec x,\eta) A_i (\vec y,
     \eta) A_j (\vec z, \eta) }  = -
     \int^{\eta_I}_{-\infty} d\eta_1 \, 2 \; \text{Im} \left[ \VEV{
     \hat H_{\rm int}(\eta_1)  \frac{\delta\phi}{M}(\vec x,
     \eta) A_i (\vec y, \eta) A_j (\vec z, \eta) }\right] .
\end{equation}
The interaction Hamiltonian is
\begin{equation}
     \hat H_{\rm int} = -\int d^3 x \,  \left(
     \frac{\eta}{\eta_I}\right)^{-2 n} \frac{\delta\phi}{M}
     \left( (A'_i)^2 - \frac{1}{2}(\partial_i A_j - \partial_j
     A_i)^2\right).
\end{equation}
Using Eqs.~(\ref{eq:dphi}) and (\ref{eq:Vi}), we find that the
expectation value on the right hand side of
Eq.~(\ref{eq:zvvtree}) is
\begin{equation}
     \VEV{\hat  H_{\rm int}(\eta')\, \frac{\delta\phi}{M}(\vec
     x,\eta_I) A_i(\vec y,\eta_I) A_j(\vec z,\eta_I)} 
     =  \int \frac{d^3k_1}{(2\pi)^3} \frac{d^3k_2}{(2\pi)^3}
     \frac{d^3k_3}{(2\pi)^3}  (2 \pi)^3 \delta(\vec k_1 +
     \vec k_2 + \vec k_3)  e^{i\vec k_1\cdot \vec x+i\vec
     k_2\cdot \vec y+i\vec k_3\cdot \vec z} (K^{(1)}_{ij} +
     K^{(2)}_{ij} ).
\label{eqn:Aexp}
\end{equation}
The functions $K^{(1)}_{ij}$ and $K^{(2)}_{ij}$ are defined as 
\begin{eqnarray}
     K^{(1)}_{ij} &=&  -\frac{2}{M^2} \delta_{ij} \left(
     \frac{\eta'}{\eta_I}\right)^{-2n} 
     \delta\phi_{k_1}(\eta')\delta\phi^*_{k_1}(\eta_I)
     \left(\frac{d}{d\eta'}A_{k_2}(\eta') \right) A^*_{k_2}(\eta_I) 
     \left(\frac{d}{d\eta'}A_{k_3}(\eta') \right) A^*_{k_3}(\eta_I),
\label{eqn:k1} \\
     K^{(2)}_{ij} &=&  -\frac{2}{M^2} (\vec k_2 \cdot \vec k_3
     \delta_{ij} - k_{2j} k_{3i}) \left(
     \frac{\eta'}{\eta_I}\right)^{-2n}
     \delta\phi_{k_1}(\eta')\delta\phi^*_{k_1}(\eta_I)
     A_{k_2}(\eta') A^*_{k_2}(\eta_I)  A_{k_3}(\eta') A^*_{k_3}(\eta_I),
 \label{eqn:k2}
\end{eqnarray}
where we indicate the scalar mode functions of the vector
potential as $A_k(\eta) = v_k(\eta)/I(\eta)$. Plugging
Eqs.~(\ref{eqn:k1})--(\ref{eqn:k2}) into Eq.~(\ref{eqn:Aexp}),
we find
\begin{equation}
     \VEV{ \frac{\delta\phi}{M}(\vec x,\eta_I) A_i(\vec
     y,\eta_I)  A_j(\vec z,\eta_I)} 
     = \int \prod_{i=1}^3 \frac{d^3 k_i}{(2 \pi)^3} {\rm
     e}^{i(\vec k_1\cdot\vec x+\vec k_2\cdot\vec y+\vec
     k_3\cdot\vec z)} (2 \pi)^3 \delta(\vec k_1 + \vec k_2 +
     \vec k_3) U_{ij},
\end{equation}
\begin{equation}
     U_{ij} = -2 \, {\rm Im} \int d\eta' \, \left(
     K^{(1)}_{ij} +K^{(2)}_{ij}   \right) = -
     \frac{\pi^2}{8}\left(\frac{H_I}{M}\right)^2\frac{1}{k_1^4}
     \left(  \delta_{ij} {\cal I}_1 + (\hat k_2 \cdot \hat k_3
     \delta_{ij} - \hat k_{2j} \hat k_{3i}){\cal I}_2\right),
\end{equation}
where we introduce the integrals
\begin{eqnarray}\label{eq:I1}
      {\cal I}_1 &=& {\rm Im}\int_1^\infty d\mu \, u_1 (i+\mu
      u_1)(-i+u_1) e^{i u_1(\mu-1)} \mu^{-2n} \cr
      &\times &
       \frac{d}{d\mu}\left[
       \mu^{\frac{1}{2}+n}H^{(1)}_{\frac{1}{2}+n}(\mu
       u_2)\right] H^{(2)}_{\frac{1}{2}+n}(u_2)
       \frac{d}{d\mu}\left[
       \mu^{\frac{1}{2}+n}H^{(1)}_{\frac{1}{2}+n}(\mu
       u_3)\right] H^{(2)}_{\frac{1}{2}+n}(u_3), \\
       {\cal I}_2 &=& {\rm Im}\int_1^\infty d\mu \, u_1 (i+\mu
       u_1)(-i+u_1) e^{i u_1(\mu-1)} u_2 u_3 \mu \cr
       & \times & H^{(1)}_{\frac{1}{2}+n}(\mu u_2) H^{(2)}_{\frac{1}{2}+n}(u_2) 
       H^{(1)}_{\frac{1}{2}+n}(\mu u_3)
       H^{(2)}_{\frac{1}{2}+n}(u_3). 
  \label{eq:I2}
\end{eqnarray}
While ${\cal I}_2$ and the magnetic-field power spectrum are
both invariant under $n \to 1+n$, ${\cal I}_1$ is not.  This is
not surprising since the interaction Hamiltonian is not
invariant under this operation. In the above, we have defined
$\mu = \eta/\eta_I$ and $u_i = -k_i \eta_I$ for $i=1,2,3$. 

The three-point correlation function for the scalar field with
the magnetic field is obtained from
\begin{eqnarray}
    \VEV{ \frac{\delta\phi}{M}(\vec x,\eta_I) \vec B(\vec
    y,\eta_I) \cdot \vec B(\vec z,\eta_I)}
    &=& -2\, {\rm Im}\,\int_{-\infty}^{\eta_I} d\eta \, \VEV{
    H_{\rm int}(\eta)\,  \frac{\delta\phi}{M}(\vec x,\eta_I)
    \vec B(\vec y,\eta_I) \cdot \vec B(\vec z,\eta_I)} \\ 
     &=& \frac{1}{a(\eta_I)^4} \left(\delta_{ij}
     \frac{\partial^2}{\partial y^k \partial z^k} -
     \frac{\partial^2}{\partial y^j \partial z^i}\right)\VEV{
     \frac{\delta\phi}{M}(\vec x,\eta) A_i(\vec y,\eta)A_j(\vec
     z,\eta)} .
\end{eqnarray}
After some calculations, the final result is
\begin{equation}
\label{eq:3pt}
     \VEV{ \frac{\delta\phi}{M}(\vec x,\eta_I) \vec B(\vec
     y,\eta_I) \cdot \vec B(\vec z,\eta_I)}
     = \int  \frac{d^3 k_1}{(2 \pi)^3} \frac{d^3 k_2}{(2
     \pi)^3} \frac{d^3 k_3}{(2 \pi)^3}  e^{i(\vec k_1\cdot\vec x+\vec k_2\cdot\vec y+\vec
     k_3\cdot\vec z)} (2 \pi)^3 \delta(\vec k_1 + \vec k_2 +
     \vec k_3) P_{3}(k_1,k_2,k_3),
\end{equation}
where
\begin{equation}
     P_{3}(k_1,k_2,k_3) = \frac{\pi^2}{8}\left(\frac{H_I}{M}\right)^2
     \frac{1}{a(\eta_I)^4}  \frac{k_2 k_3}{k_1^4} \left(2 \hat
     k_2 \cdot \hat k_3\, {\cal I}_1 + (1+(\hat k_2\cdot\hat
     k_3)^2) \,{\cal I}_2\right).
\label{eq:p3}
\end{equation}
\end{widetext}
From statistical isotropy, the function $P_3(k_1,k_2,k_3)$
depends only on the magnitudes of the three wavevectors, and we
have used $\hat k_2 \cdot \hat k_3 = (k_1^2-k_2^2-k_3^2)/(2k_2
k_3)$ in Eq.~(\ref{eq:p3}).  Eqs.~(\ref{eq:3pt}) and
(\ref{eq:p3}), along with Eqs.~({\ref{eq:I1})--(\ref{eq:I2}),
form the main results on which our subsequent analysis is
based.

\section{Analysis of Cross-Correlation}
\label{sec:analysis}

We would like to analyze the cross-correlation between the
primordial magnetic field and the scalar field to determine if
there is any imprint or unique signature that would indicate the
scalar field's role in the amplification. 
 
\subsection{The Amplified Cross-Correlation Power Spectrum}

To start, we calculate the cross-correlation power spectrum for
several trial cases. The integrals ${\cal I}_{1,2}$ can be
evaluated analytically for integer values of $n$. In most cases,
the results are cumbersome, so we assume $u_i \ll 1$ after
carrying out the integrals in order to shorten the expressions.
For example, for $n=0$,
\begin{equation}
     {\cal I}_1 = -{\cal I}_2 =  -\frac{4}{\pi^2} \frac{u_1(u_1
     + \omega)}{\omega^2},
\end{equation}
where $\omega = u_1 + u_2 + u_3$. Plugging in these results, we
find
\begin{equation}
     P_3(k_1,k_2,k_3) = \frac{(H_I/M)^2}{a(\eta_I)^4} \frac{(2 k_1 + k_2 +
     k_3)(k_1-k_2-k_3)^2}{8 k_1^3 k_2 k_3 },
\end{equation}
where we have used $\cos\theta = \hat k_2 \cdot \hat k_3 =
(k_1^2 - k_2^2 - k_3^2)/(2 k_2 k_3)$ for the angle between the
vectors $k_2$ and $k_3$.

Amplification occurs for $n>0$ and $n<-1$, so that for comparison
we consider integer cases $n=1$ and $n=-2$ whereupon the
integration simplifies.  For $n=1$, we find
\begin{eqnarray}
     {\cal I}_1 &=& \frac{4}{\pi^2}\frac{u_1(u_1 + \omega)}{u_2
     u_3 \omega^2}, \\
     {\cal I}_2 &=& \frac{4}{\pi^2} \frac{u_1(\omega^3 + u_1^2
     \omega -u_1 \omega^2 - u_2 u_3 \omega - u_1 u_2 u_3)}{u_2^2
     u_3^2 \omega^2}.\nonumber \\
\end{eqnarray}
For $n=-2$,
\begin{eqnarray}
     {\cal I}_1 &=& -\frac{4}{\pi^2} \frac{u_1}{u_2^3 u_3^3
     \omega^2} \left( 3u_1^3 \omega^2(\gamma + \ln\omega)  + 3
     u_1^3 u_2 u_3 \right. \nonumber \\
     & &  \left. + 3 u_1^2(u_2^2 u_3 + u_2 u_3^2 - \omega^3)
     -u_2^2 u_3^2(\omega+u_1)\right),\\ \cr
     {\cal I}_2 &=& \frac{4}{\pi^2} \frac{u_1(\omega^3 + u_1^2
     \omega -u_1 \omega^2 - u_2 u_3 \omega - u_1 u_2 u_3)}{u_2^2
     u_3^2 \omega^2}.\nonumber \\
\end{eqnarray}
In the case of most interest, $n=2$, the integrals yield
\begin{eqnarray}
     {\cal I}_1 &\simeq & \frac{36}{\pi^2}\frac{u_1}{u_2^3 u_3^3
     \omega^2} \left(\omega^3 - u_1 u_2 u_3 - \omega(u_1 u_2 +
     u_1 u_3 + u_2 u_3) \right), \nonumber \\ \\
     {\cal I}_2 &\simeq& \frac{36}{\pi^2}\frac{u_1}{u_2^4 u_3^4
     \omega^2} \left(-3 u_1^3 \omega^2(\gamma + \ln\omega) +
     \omega^5 - 3 u_1 \omega^4 \right. \nonumber \\
     & & + 3(2 u_1^2 - u_2 u_3) \omega^3
     + (3 u_1 u_2 u_3 - u_1^3)\omega^2 \nonumber \\
     & &  \left. + (u_2^2 u_3^2 - 3
     u_1^2 u_2 u_3)\omega + u_1 u_2^2 u_3^2\right).
\end{eqnarray}
Since $|k \eta_I| \ll 1$ we have discarded subdominant terms
from the above results. We note that the $\ln\omega$ term above
results in a large numerical coefficient, since $-k \eta_I \sim
10^{-27}$ for modes that are just entering the horizon today.
These expressions are inserted into Eq.~(\ref{eq:p3}) to find
the cross-correlation power spectrum.

\subsection{The Real-Space Cross-Correlation Coefficient}

Our next step is to determine the dimensionless magnitude of the
cross-correlation; i.e., how strongly does the magnetic-field
energy density correlate with the scalar-field perturbation?  We
thus now calculate the zero-lag cross-correlation $\VEV{ (\delta
\phi) B^2}$ in units of $\VEV{(\delta\phi)^2}^{1/2} \VEV{B^2}$.

This cross-correlation amplitude, evaluated in the coincidence
limit, can be evaluated as follows.  Starting from
Eqs.~(\ref{eq:3pt})--(\ref{eq:p3}), we evaluate the $\vec k_1$
integration to eliminate the delta function. The remaining
integrand depends only on the magnitudes $k_2$, $k_3$, and
$\theta$, the angle between the two vectors:
\begin{equation}
     \VEV{ (\delta\phi) B^2 }  = \frac{M}{8 \pi^4} \int
     k_2^2 dk_2 \, k_3^2 dk_3 \, d(\cos\theta) \, P_3(k_1, k_2,
     k_3)
     \label{eqn:p3integral}
\end{equation}
where $k_1 = (k_2^2 + k_3^2 + 2 k_2 k_3
\cos\theta)^{1/2}$. However, we can replace the
$\theta$ integral by $k_1$, whereby
\begin{equation}
     \VEV{ (\delta\phi) B^2}  = \frac{M}{8 \pi^4} \int
     k_2 dk_2 \, k_3 dk_3 \int_{|k_2 - k_3|}^{k_2 + k_3} k_1
     dk_1 \, P_3(k_1, k_2, k_3).
\end{equation}
Since the integrand is invariant under the exchange of $k_2$ and
$k_3$, we can replace $P_3 \to 2 P_3 \theta(k_2 - k_3)$ and
remove the absolute-value sign from the lower limit of
integration. We implement cutoffs at both large and small $k$,
for the ultraviolet and infrared divergences that arise in both
the scalar and magnetic-field spectra. The cross-correlation for
$n=0$ and $n=2$ are
\begin{widetext}
\begin{equation}
\label{eq:cross}
     \VEV{ (\delta\phi) B^2  } \simeq \frac{M}{16 \pi^4
     a^4(\eta_I)}\left(\frac{H_I}{M}\right)^2
     \times \begin{cases} k_\text{max}^4 \left( \ln{r} -
     \frac{25}{12}\right), & n=0, \cr 
     \eta_I^{-4} \left( 100 + 24 \ln^3r - 72 \ln^2r
     \ln(-k_\text{max}\eta_I)\right), & n=2, \cr
\end{cases}
\end{equation}
where $r=k_\text{max}/k_\text{min}$, and $k_\text{max}$ and
$k_\text{min}$ are upper and lower bounds on the run of
wavevectors. In practice, we expect to link the minimum
wavevector with the Hubble scale, $k_\text{min} \simeq 2 \pi
H_0$, and the maximum wavevector with some galactic scale,
$k_\text{max} \simeq 2 \pi/\lambda$ where $\lambda\sim$kpc.
Since $|k \eta_I |\ll 1$, we have discarded subdominant terms
from the above results. The dimensionless cross-correlation
coefficient $X_{\delta\phi B^2}$, formed from the ratio of the
cross-correlation with the root-mean-square amplitudes of the
scalar and magnetic fields gives
\begin{equation}
\label{eq:realcross}
     X_{\delta\phi B^2} \equiv \frac{\VEV{ \delta\phi
     B^2 } }{ \left(\delta\phi \right)_\text{rms} B_\text{rms}^2   } \simeq
     \begin{cases}
     \frac{1}{\pi}\left( \frac{H_I}{M}\right)\left({\ln r -
     \frac{25}{12}}\right)/{\sqrt{\ln r}}, & n=0, \cr 
     \frac{4}{9\pi}\left( \frac{H_I}{M}\right) \left(25 + 6
     \ln^3 r - 18 \ln^2 r
     \ln(-k_\text{max}\eta_I)\right)/{\sqrt{\ln^3 r}}, & n=2. \cr
\end{cases}
\end{equation}
\end{widetext}
Considering a sufficiently wide range of scales, e.g. $r
\gtrsim 10^4$, then $X(n=0)\simeq (H_I/M)\sqrt{\ln r}/\pi $ and
$X(n=2) \simeq 8(H_I/M)\sqrt{\ln r}  \ln(-k_\text{max}\eta_I)
/\pi$. Using $-k_\text{max}\eta_I \sim 10^{-27}$ then the
cross-correlation coefficient in the presence of the
amplification mechanism is enhanced by a factor of $\sim 500$
over the case without the magnetic-field amplification
mechanism.  When the full range of inflationary length scales is
taken, $r \sim 10^{27}$, then $X(n=2) \simeq 2 \times 10^3
(H_I/M)$. Since the cross-correlation coefficient cannot exceed
unity, we infer an upper bound of $H_I/M \lesssim 5 \times
10^{-4}$ which is consistent with naive expectations based on an
inflationary scenario.

\subsection{The Behavior in Fourier-Space}

We now evaluate the triangle-shape dependence of the full
three-point correlation function in Fourier space.  To do so, we
evaluate a ratio of the form,
\begin{equation}
     \frac{P_3(k_1,k_2,k_3)}{\sqrt{P_{\delta\phi}(k_1) P_B(k_2)
     P_B(k_3)}},
\end{equation}
to normalize the cross-correlation power spectrum. However,
since this ratio is not dimensionless, given our Fourier
conventions, we go to a discretized Fourier transform,
\begin{equation}
     \int \frac{d^3k}{(2 \pi)^3} \to \frac{1}{V}\sum_{\vec n},
\end{equation}
and likewise replacing the Dirac delta function with a Kronecker
delta,
\begin{equation}
     (2 \pi)^3 \delta(\vec k_1+ \vec k_2) \to V \delta_{\vec n_1,\vec n_2}.
\end{equation}
We presume a maximum length, $L$, so that the volume is $V=L^3$
and mode numbers are $k_i = 2 \pi n_i/L$. The scalar-field and
magnetic-field power spectra are now
\begin{eqnarray}
     \VEV{ (\delta\phi/M)^2 } &=& \sum_{\vec n}
     e^{i\vec n \cdot (\vec x-\vec
     y)/L } \widetilde
     P_{\delta\phi},\\ 
     \widetilde P_{\delta\phi} &=& V^{-1} P_{\delta\phi}/M^2, \\
     \VEV{ B^2} &=& \sum_{\vec n} e^{i\vec n \cdot( \vec x - \vec y)/L }
     \delta_{\vec n_1,\vec n_2} \widetilde P_{B},\\
     \widetilde P_{B} &=& V^{-1} P_{B},
\end{eqnarray}
so that $\widetilde P_{\delta\phi}$ is dimensionless and
$\widetilde P_{B}$ has units of (energy)${}^4$. The three-point
function becomes
\begin{eqnarray}
     \VEV{ \frac{\delta\phi}{M} B^2} &=& \sum_{\vec n_1+\vec
     n_2+\vec n_3=0} e^{i(\vec n_1 \cdot
     \vec x + \vec n_2\cdot\vec y+ \vec n_3\cdot\vec z)/L }
     \widetilde P_{3},\nonumber\\ 
     \widetilde P_{3} &=& V^{-2} P_{3},
\end{eqnarray}
where $\widetilde P_3$ has units of (energy)${}^4$. We can now
build a dimensionless cross-correlation coefficient,
\begin{equation}
     C_n = \frac{\widetilde P_{3}(n_1,n_2,n_3)}{\sqrt{\widetilde
     P_{\delta\phi}(n_1) \widetilde P_B(n_2) \widetilde
     P_B(n_3)}},
\end{equation}
where $n_i$ for $i=1,2,3$ are the magnitudes of vectors $\vec
n_i$ that form a closed triangle.

\begin{figure*}[htbp]
\vspace{20pt}
\begin{center}
\includegraphics[scale=0.66]{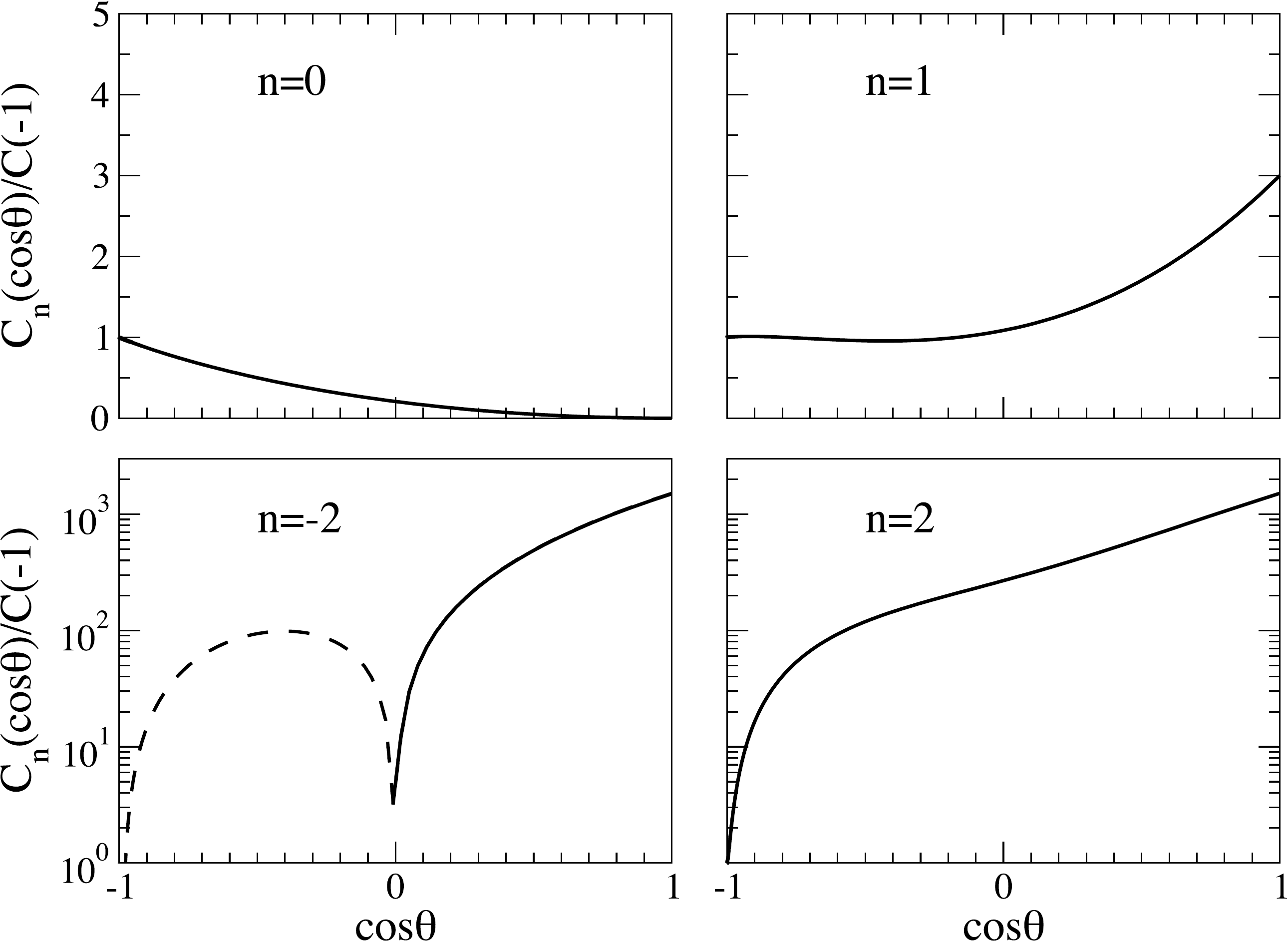}
\end{center}
\caption{The ratio $C_n(\cos\theta)/C_n(-1)$ is shown for $n=0$,
    $1$, $2$, and $-2$, as functions of $\cos\theta$. In the
    $n=-2$ panel, the dashed line indicates where the absolute
    value has been taken. In the $n=2,\, -2$ cases  we have used
    $2 \pi n_1 |\eta_I/L| \sim  10^{-27}$ corresponding
    approximately to a Gpc length scale. Note that the case of
    cosmological interest is $n=2$.}
\label{fig:fig1}
\end{figure*}

\begin{figure*}[htbp]
\vspace{20pt}
\begin{center}
\includegraphics[scale=0.66]{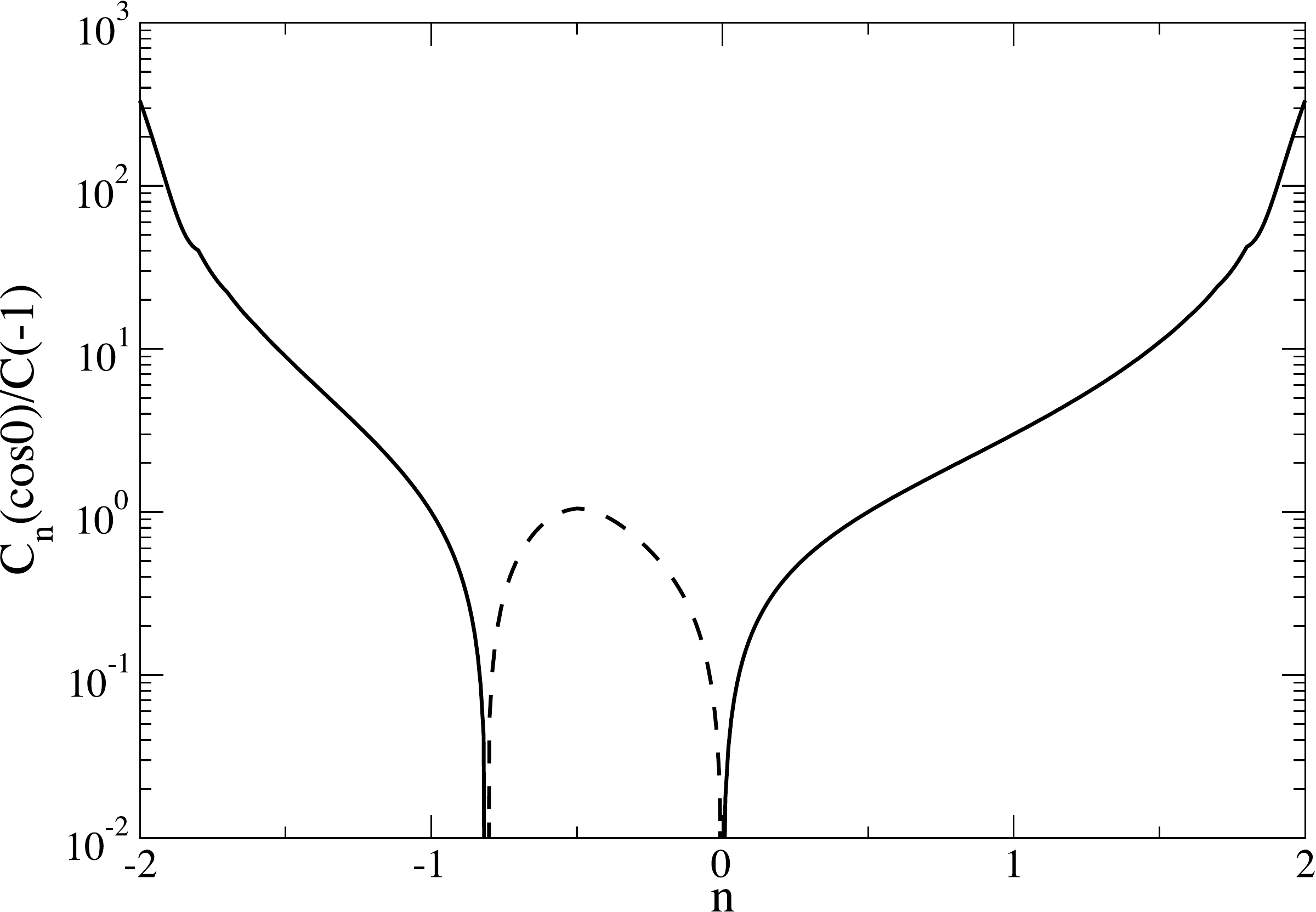}
\end{center}
\caption{The ratio $C_n(1) / C(-1)$, the ratio of the
     discretized Fourier-space cross-correlation coefficients
     for the flattened triangle to that of the universal result
     for the squeezed triangle, is shown as a function of $n$.
     No amplification, $n=0$, yields zero cross-correlation.
     Hence, the flattened triangle may be used as an indicator
     of an amplification mechanism. The ratio is negative along
     the dashed line, where we have taken the absolute value. We
     have set $2\pi n_1 |\eta_I/L| \sim 10^{-6}$ for ease  of
     numerical computation; using $2\pi n_1 |\eta_I/L| \sim
     10^{-27}$ to represent Gpc scales boosts the curve up to
     $10^3$ near $n=\pm 2$. Note that the case of cosmological
     interest is $n=2$.}
\label{fig:fig2}
\end{figure*}

For isosceles triangles with $n_2 = n_3$, the correlation $C_n$
obtained for the case $n=0$ and $n=1$ is
\begin{eqnarray}
     C_0 &=& \frac{1}{8 \pi^{3/2}} \frac{H_I}{M}
     \frac{(n_1+n_2)(n_1 - 2 n_2)^2}{n_1^{3/2} n_2^3}, \nonumber
     \\
     C_{1} &=& \frac{1}{16 \pi^{3/2}} \frac{H_I}{M}
     \frac{N}{(n_1 + 2 n_2)
     n_1^{3/2} n_2^5 },\nonumber \\
     N&=& n_1^6 + 2 n_1^5 n_2 - 2 n_1^4 n_2^2 - 6 n_1^3 n_2^3
     \nonumber \\
     & & +      4 n_1^2 n_2^4 + 8 n_1 n_2^5 + 16 n_2^6,
\end{eqnarray}
where $-1 \le \cos\theta = \frac{1}{2}\frac{n_1^2}{n_2^2}-1 \le
1$. An expression for $C_2$ is easily calculated, but the result
is rather long and unenlightening. The behavior of
$C_n(\cos\theta)$ for $n=0,\,1,\,2$, and $-2$ is illustrated in
Fig.~\ref{fig:fig1}.

We find that there are two interesting limits for isosceles triangles with
$n_2 = n_3$,  first a squeezed triangle, with $1 \le n_1 \ll
n_2$ or $\theta = \pi$, and second a flattened triangle, with
$n_2 = n_1/2$ or $\theta=0$.  For the squeezed triangle we  find
the universal result,
\begin{equation}
     C_n(\cos\pi) =  \frac{\sqrt{2}}{(2 \pi n_1)^{3/2}}
     \frac{H_I}{M},
     \label{eqn:universal}
\end{equation}
for all values of $n$, as borne out by numerical integration for
non-integer $n$. We suspect that this triangle configuration,
with small $n_1$ and large $n_2, n_3$, dominates the integration
in Eq.~(\ref{eq:realcross}), as a way to help explain the
similarities seen in the real-space cross-correlation
coefficients for different values of the index $n$.

The result, Eq.~(\ref{eqn:universal}), suggests a natural reference
point, so that a general expression for the discretized
Fourier-space dependence of the cross-correlation is
\begin{equation}
     C_n(\cos\theta)/C_n(\cos\pi) = \frac{\pi}{4 \epsilon
     n_1}\frac{2 \cos\theta {\cal I}_1 + (1 + \cos^2\theta){\cal
     I}_2}{|H^{(1)}_{1/2+n}(\epsilon
     n_2)H^{(1)}_{1/2+n}(\epsilon n_3)|},
\end{equation}
where $\epsilon \equiv -2 \pi \eta_I/L \ll 1$. 

For a flattened triangle, we have $C_0(\cos 0)/C_n(\cos\pi) =0$,
$C_{1}(\cos 0)/C_n(\cos\pi)=3$, and $C_{-2}(\cos 0)/C_n(\cos\pi)
=  12 \left( 2- \gamma - \ln(2\epsilon)\right)$, where $\gamma$
is the Euler-Mascheroni constant.  Note that the
cross-correlation vanishes for the unamplified case ($n=0$), but
grows large for $n=-2$, where the argument of the log is $\sim
10^{-27}$ for modes entering the horizon today. The behavior of
$C_n(\cos 0)/C_n(\cos\pi)$ as a function of $n$ is shown in
Fig.~\ref{fig:fig2}.

To show the full Fourier-space triangle dependence of the
cross-correlation, we define the quantity
\begin{equation}
R \equiv  \left(\frac{n_2}{n_3}\right)^{2} \frac{C_n(\cos\theta)}{C_n(\cos\pi)}
\end{equation}
and introduce the variables $x_{23} \equiv n_2/n_3$ and $x_{13}
\equiv n_1/n_3$, where $0\le x_{23} \le 1$ and $1-x_{23} \le
x_{13} \le 1+x_{23}$ covers the full set of triangles. The
behavior for the cases $n=0,\,1,\,2$, and $-2$ are shown in
Fig.~\ref{fig:fig3}. The Figure helps illustrate the difference
between the amplified and unamplified ($n=0$) cases, and  also
shows that the maximum value of $R$ in the amplified case occurs
for the flattened triangles, corresponding to the line $x_{13} =
1+x_{23}$, along which $\theta = 0$.  Squeezed triangles, where
$\theta = \pi$, are located along $x_{13} = 1-x_{23}$.

\begin{figure*}[htbp]
\vspace{20pt}
\begin{center}
\includegraphics[scale=0.6]{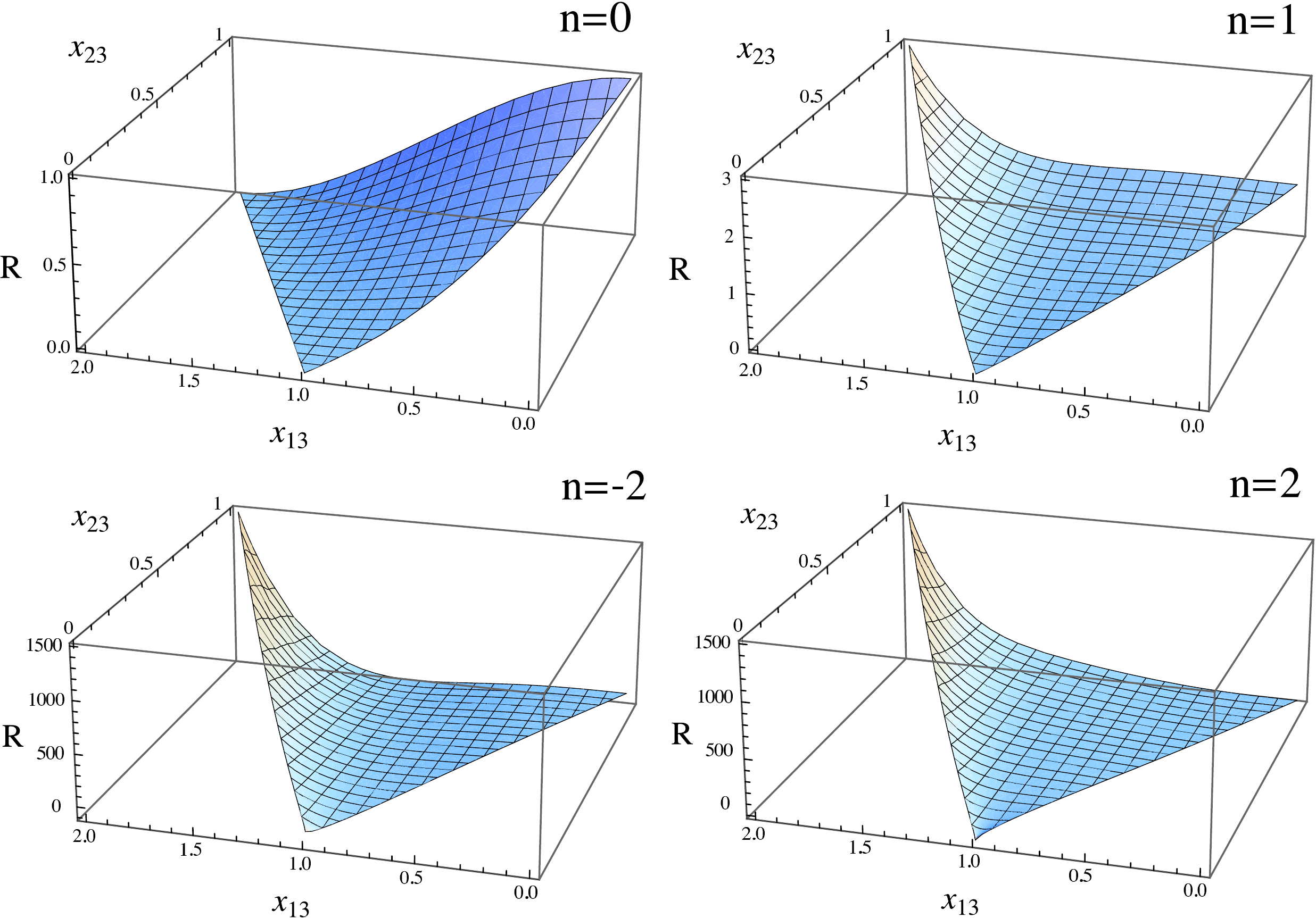}
\end{center}
\caption{The quantity $R$, defined in the text as the ratio of
     the Fourier-space cross-correlation coefficient to that of
     the universal result for the squeezed triangle, times a
     factor $x_{23}^2$, is shown as a function of the triangle
     side lengths.   We have set  $2\pi n_3 |\eta_I/L| \sim
     10^{-27}$ for the cases $n=\pm 2$. Note that the case of
     cosmological interest is $n=2$.}
\label{fig:fig3}
\end{figure*}

The amplitude of the zero-lag, real-space cross-correlation is
dominated by the cut-offs to the range of Fourier modes. At
finite Fourier mode, however, the flattened isosceles triangle
produces the largest cross-correlation amplitude and hence the
best opportunity to observe the signature of the mechanism
responsible for amplifying the primordial magnetic field.

\section{\label{sec:conclusion}Conclusions}

We have considered a toy model in which a scalar field is
coupled to electromagnetism in a fixed de Sitter background. The
homogeneous time evolution of the scalar field breaks the
conformal invariance of electromagnetism resulting in quantum
production of magnetic fields in addition to quantum production
of scalar-field fluctuations.  We then calculated the
cross-correlation between the scalar field and the
magnetic-field energy density.  The dimensionless
cross-correlation coefficient  is proportional to the ratio
$H_I/M$, which must be small if the effect of the scalar-field
perturbation on the electromagnetic part of the Lagrangian can
be considered small.  However, this small quantity may be
multiplied by a numerically large ($\lesssim500$) coefficient
suggesting a possibly strong (even order-unity)
cross-correlation.

We also studied the full triangle-shape dependence of the
three-point correlation function in Fourier space.  We find that
it is nonzero for squeezed triangles (wherein the short Fourier
component is that associated with the scalar-field mode), but
may be considerably larger for flattened triangles (where the
long Fourier mode is associated with the scalar field and twice
as long as those associated with the magnetic field).  These
shape dependences may be useful if such correlations are to be
sought in the data.

Although we treat it as a toy model, our calculation provides
the correlation between the curvaton and magnetic fields if the
scalar field is identified as the curvaton.  If primordial
perturbations are further due to curvaton fluctuations, then the
scalar-field--magnetic-field cross-correlations derived here
describe density-perturbation--magnetic-field correlations in
the Universe today.  If the scalar field is the inflaton, then
there are additional steps to relate the scalar-field
perturbation to the density-perturbation amplitude in the
Universe today \cite{inprogress}.

The cross-correlation between primordial-seeded density
perturbations and magnetic fields may be amenable to detection
through the cosmic microwave background (CMB). Cosmic magnetic
fields present during the recombination era contribute to the
CMB temperature and polarization signals. (See
Ref.~\cite{Lewis:2004kg} for a detailed study.) Magnetic fields
along the line of sight further distort the CMB by converting
E-mode polarization into B-mode polarization, through Faraday
rotation
\cite{Kosowsky:1996yc,Kosowsky:2004zh,Giovannini:2008aa,Kahniashvili:2008hx,Kristiansen:2008tx}.
Primordial magnetic fields may also leave a non-Gaussian imprint
on the statistics of the anisotropy pattern
\cite{Seshadri:2009sy,Caprini:2009vk,Cai:2010uw,Shiraishi:2010kd,Brown:2010jd}.
Current observations set the upper bound on a primordial
magnetic field at the nG level
\cite{Giovannini:2009zq,Yamazaki:2010nf,Kahniashvili:2010wm}.
(We also note that there have been claims of a lower bound on an
extragalactic field  \cite{Neronov:1900zz}.)

The correlation may also be accessible through a combined survey
of large-scale structure and Faraday rotation 
\cite{Stasyszyn:2010kp}. The proposed SKA telescope, which is
projected to be sensitive to variations of $0.1 \, \text{nG}$
across 100 Mpc, and LOFAR, which aims to explore the nG fields
in intergalactic
media~\cite{Rottgering:2006ms,Beck:2007,Beck:2009ew,Gaensler:2009ka},
may offer more direct means to probe for the cross correlation.
The detectability of the effect studied here will be the subject
of future work \cite{inprogress}.

\acknowledgments
MK thanks the support of the Miller Institute for Basic Research
in Science and the hospitality of the Department of Physics at
the University of California, Berkeley, where part of this work
was completed.  RC and LM thank Caltech for hospitality, where
part of this work was completed.  This work was supported in
part by NSF AST-0349213 at Dartmouth College and by DOE
DE-FG03-92-ER40701, NASA NNX10AD04G, and the Gordon and Betty
Moore Foundation at Caltech. 

\vfill  



\begin{thebibliography}{}

\bibitem{Weinberg:2008hq}
  S.~Weinberg,
  Phys.\ Rev.\  {\bf D77}, 123541 (2008).
  [arXiv:0804.4291 [hep-th]].
  
\bibitem{Senatore:2010wk}
  L.~Senatore, M.~Zaldarriaga,
  [arXiv:1009.2093 [hep-th]].
  
\bibitem{Bartolo:2004if}
  N.~Bartolo, E.~Komatsu, S.~Matarrese, A.~Riotto,
  Phys.\ Rept.\  {\bf 402}, 103-266 (2004).
  [astro-ph/0406398].
  
\bibitem{Komatsu:2009kd}
  E.~Komatsu, N.~Afshordi, N.~Bartolo, D.~Baumann, J.~R.~Bond, E.~I.~Buchbinder, C.~T.~Byrnes, X.~Chen {\it et al.},
  [arXiv:0902.4759 [astro-ph.CO]].

\bibitem{Huterer:2010en}
  D.~Huterer, S.~Shandera, E.~Komatsu,
  [arXiv:1012.3744 [astro-ph.CO]]..
  
\bibitem{Turner:1987bw}
  M.~S.~Turner, L.~M.~Widrow,
  Phys.\ Rev.\  {\bf D37}, 2743 (1988).

\bibitem{Ratra:1991bn}
 B.~Ratra,
 Astrophys.\ J.\  {\bf 391}, L1 (1992).

\bibitem{Widrow:2002ud}
  L.~M.~Widrow,
  Rev.\ Mod.\ Phys.\  {\bf 74}, 775-823 (2002).
  [astro-ph/0207240].

\bibitem{Bamba:2003av}
  K.~Bamba, J.~Yokoyama,
  Phys.\ Rev.\  {\bf D69}, 043507 (2004).
  [astro-ph/0310824].
 
\bibitem{Kunze:2007ph}
  K.~E.~Kunze,
  Phys.\ Rev.\  {\bf D77}, 023530 (2008).
  [arXiv:0710.2435 [astro-ph]].
  
\bibitem{Campanelli:2008qp}
  L.~Campanelli, P.~Cea, G.~L.~Fogli, L.~Tedesco,
  Phys.\ Rev.\  {\bf D77}, 123002 (2008).
  [arXiv:0802.2630 [astro-ph]].

\bibitem{Demozzi:2009fu}
  V.~Demozzi, V.~Mukhanov, H.~Rubinstein,
  JCAP {\bf 0908}, 025 (2009).
  [arXiv:0907.1030 [astro-ph.CO]].

\bibitem{Kunze:2009bs}
  K.~E.~Kunze,
  Phys.\ Rev.\  {\bf D81}, 043526 (2010).
  [arXiv:0911.1101 [astro-ph.CO]].

\bibitem{Kandus:2010nw}
  A.~Kandus, K.~E.~Kunze and C.~G.~Tsagas,
  Phys.\ Rept.\  {\bf 505}, 1 (2011)
  [arXiv:1007.3891 [astro-ph.CO]].

\bibitem{Lewis:2004kg}
  A.~Lewis,
  Phys.\ Rev.\  {\bf D70}, 043518 (2004).
  [astro-ph/0403583].

\bibitem{Kosowsky:1996yc}
  A.~Kosowsky, A.~Loeb,
  Astrophys.\ J.\  {\bf 469}, 1-6 (1996).
  [astro-ph/9601055].
  
\bibitem{Kosowsky:2004zh}
  A.~Kosowsky, T.~Kahniashvili, G.~Lavrelashvili, B.~Ratra,
  Phys.\ Rev.\  {\bf D71}, 043006 (2005).
  [astro-ph/0409767].

\bibitem{Giovannini:2008aa}
  M.~Giovannini, K.~E.~Kunze,
  Phys.\ Rev.\  {\bf D78}, 023010 (2008).
  [arXiv:0804.3380 [astro-ph]].
  
\bibitem{Kahniashvili:2008hx}
  T.~Kahniashvili, Y.~Maravin, A.~Kosowsky,
  Phys.\ Rev.\  {\bf D80}, 023009 (2009).
  [arXiv:0806.1876 [astro-ph]].
  
\bibitem{Kristiansen:2008tx}
  J.~R.~Kristiansen and P.~G.~Ferreira,
  Phys.\ Rev.\  D {\bf 77}, 123004 (2008)
  [arXiv:0803.3210 [astro-ph]].
    
\bibitem{Seshadri:2009sy}
  T.~R.~Seshadri and K.~Subramanian,
  Phys.\ Rev.\ Lett.\  {\bf 103}, 081303 (2009)
  [arXiv:0902.4066 [astro-ph.CO]].
  
\bibitem{Caprini:2009vk}
  C.~Caprini, F.~Finelli, D.~Paoletti, A.~Riotto,
  JCAP {\bf 0906}, 021 (2009).
  [arXiv:0903.1420 [astro-ph.CO]].
    
\bibitem{Cai:2010uw}
  R.~G.~Cai, B.~Hu and H.~B.~Zhang,
  JCAP {\bf 1008}, 025 (2010)
  [arXiv:1006.2985 [astro-ph.CO]].
  
\bibitem{Shiraishi:2010kd}
  M.~Shiraishi, D.~Nitta, S.~Yokoyama, K.~Ichiki and K.~Takahashi,
  Prog.\ Theor.\ Phys.\  {\bf 125}, 795 (2011)
  [arXiv:1012.1079 [astro-ph.CO]].
  
\bibitem{Brown:2010jd}
  I.~A.~Brown,
  Astrophys.\ J.\  {\bf 733}, 83 (2011)
  [arXiv:1012.2892 [astro-ph.CO]].
  
\bibitem{Giovannini:2009zq}
  M.~Giovannini,
  Phys.\ Rev.\  D {\bf 79}, 121302 (2009)
  [arXiv:0902.4353 [astro-ph.CO]].
   
\bibitem{Yamazaki:2010nf}
  D.~G.~Yamazaki, K.~Ichiki, T.~Kajino and G.~J.~Mathews,
  Phys.\ Rev.\  D {\bf 81}, 023008 (2010)
  [arXiv:1001.2012 [astro-ph.CO]].

\bibitem{Kahniashvili:2010wm}
  T.~Kahniashvili, A.~G.~Tevzadze, S.~K.~Sethi, K.~Pandey, B.~Ratra,
  Phys.\ Rev.\  {\bf D82}, 083005 (2010).
  [arXiv:1009.2094 [astro-ph.CO]].

\bibitem{Weinberg:2005vy}
  S.~Weinberg,
  Phys.\ Rev.\  {\bf D72}, 043514 (2005).
  [hep-th/0506236].

\bibitem{Neronov:1900zz}
  A.~Neronov, I.~Vovk,
  Science {\bf 328}, 73-75 (2010).
  [arXiv:1006.3504 [astro-ph.HE]].

\bibitem{Stasyszyn:2010kp}
  F.~Stasyszyn, S.~E.~Nuza, K.~Dolag, R.~Beck, J.~Donnert,
  [arXiv:1003.5085 [astro-ph.CO]].

\bibitem{Rottgering:2006ms}
  H.~J.~A.~Rottgering, R.~Braun, P.~D.~Barthel, M.~P.~van Haarlem, G.~K.~Miley, R.~Morganti, I.~Snellen,
  [astro-ph/0610596]. 
 
 \bibitem{Beck:2007}
 R. Beck,
 Adv. Radio Sci., {\bf 5}, 399-405, (2007);
 [www.adv-radio-sci.net/5/399/2007/].
 
\bibitem{Beck:2009ew}
  R.~Beck,
  [arXiv:0912.2918 [astro-ph.IM]].
 
\bibitem{Gaensler:2009ka}
  B.~M.~Gaensler,
  [arXiv:0901.2952 [astro-ph.IM]].

\bibitem{inprogress}
   L.~Motta, R.~R.~Caldwell, and M.~Kamionkowski, {\it in progress}.
   
     
\end{thebibliography}
\end{document}